# Pulsejet engine dynamics in vertical motion using momentum conservation

Tiberius O. Cheche

University of Bucharest, Faculty of Physics, Bucharest 077125, Romania

E-mail: cheche@gate.sinica.edu.tw

**Abstract.** The momentum conservation law is applied to analyse the dynamics of pulsejet engine in vertical motion in a uniform gravitational field in the absence of friction. The model predicts existence of a terminal speed given frequency of the short pulses. The conditions that the engine does not return to the starting position are identified. The number of short periodic pulses after which the engine returns to the starting position is found to be independent of the exhaust velocity and gravitational field intensity for certain frequency of the pulses. The pulsejet engine and turbojet engine aircraft models of dynamics are compared. Also the octopus dynamics is modelled. The paper is addressed to intermediate undergraduate students of classical mechanics and aerospace engineering.

## 1. Introduction

Systems generating reaction thrust by periodic fluid intake/exhaust cycle is common in nature and technology, for example the motion of some cephalopods [1, 2] and the pulsejet-powered aircrafts [3-5]. The basic processes occurring in these types of engines are well understood. The engine works by consumption of energy in an intake/exhaust cycle in which a fluid amount is taken into the engine and then expelled from it at a high relative velocity. A simple model for the pulsejet engine dynamics can be constructed by using one of the fundamental laws of the mechanics, namely, the principle of conservation of linear momentum, and ignoring the physical characteristics of the cycle (which, for example, in the case of aircraft would involve a thermodynamic description of air compression by ignition of air-fuel mixture). As a case study, the problem offers the opportunity to understand and apply momentum conservation to the thrust engine. In the case of cephalopods the energy is supplied by the animal locomotion system [1, 2], while for vehicles equipped with pulsejet engines, the energy is supplied by fuel, which is burned [3-5]. For simplicity, we discuss the dynamics of the pulsejet engine only in the vertical direction, and use the same model to obtain an approximate description of cephalopod motion and vertical take-off and landing (VTOL) of military aircraft [6].

The pulse duration and frequency, the buoyancy, and the drag force have significant influence on the engine dynamics. In this paper, we introduce the following simplifying assumptions: (i) the gravitational field is uniform; (ii) since buoyancy acts against gravity we include its effect by reducing intensity of the gravitational field; (iii) where the engine works by fuel ignition, we ignore the added fuel which is tiny relative to the amount of air drawn in during cycle and we assume that the aircraft has constant mass; (iv) the pulse duration is much less than the time interval between successive pulses; (v) the motion is frictionless.

The structure of the paper is as follows. In section 2, we introduce the theoretical modelling emphasizing the pedagogical aspects in describing the engine kinematics. In addition to the educational introduction of the momentum conservation we report, as interesting physics, the number of pulses in the periodic frictionless motion the engine returns to the starting position is independent of the exhaust velocity and gravity for certain frequency of the pulses. In subsection 2.1, we compare our model with other two frictionless jet engine models. In section 3, we apply the model to the vertical take-off aircraft and to the vertical motion of some octopi. The results obtained in modelling the kinematics of the pulsejet and turbojet engine are compared. Also, the limits of the frictionless model in explaining the real motion are discussed. In the last section, we present conclusions.

**2. Theoretical Modelling**

At conceptual level, we start by defining the two body system formed by the engine of mass $M$ and the amount of fluid of mass $m$ from the environment. The forces between the two bodies generated in the intake and exhaust processes are internal forces of the system of action-reaction type, consequently they can not modify the system (total) momentum. The intake process by which the engine takes the fluid amount from the environment is modelled as a perfectly inelastic collision. The exhaust process by which the fluid amount is expelled from the engine is also modelled as a perfectly inelastic collision, but time-reversed. The external force that acts on the system is the gravity. The pulse is defined as the pair of fluid intake and exhaust processes, and the pulse duration as the time interval incorporating successive intake and exhaust processes. Under the short pulse approximation (assumption (iv) from Introduction), the pulse itself has negligible duration. During a short pulse, position of the engine remains unchanged. Next we introduce the pulsejet engine frictionless model of dynamics.

The engine starts moving at the initial time $t=0$ from the origin of an upward directed vertical axis of the laboratory frame (LF) after the first pulse. The pulses are short and periodic with period $\tau$. To explain the dynamics of the pulsejet engine, the momentum conservation law for the two body system is applied with respect to the LF according to the scenario suggested by Figure 1. We denote the exhaust speed (with respect to the engine) by $u$ and the gravitational field intensity by $g$ ($u$ and $g$ define the magnitude of vectors **u** and **g**, respectively, in Figure 1). For the *first* pulse, we consider momentum conservation in the exhaust process and write,

$$0 = MV_0 - mu, \qquad (1)$$

to obtain the engine velocity (with respect to the LF) just after the first pulse (at $t=0_+$),

$$V_0 = fu, \qquad (2)$$

where $f = m/M$ and $V_0$ is the magnitude of vector $\mathbf{V}_0$ (the engine moves in opposite direction to the gravitational field). Due to the momentum conservation, in the LF the system momentum remains zero both just before and just after the intake (at $t=0_-$ and $t=0$, respectively). After the first pulse, we describe the motion by a 3-stage process of *freefall-intake-exhaust* as follows.

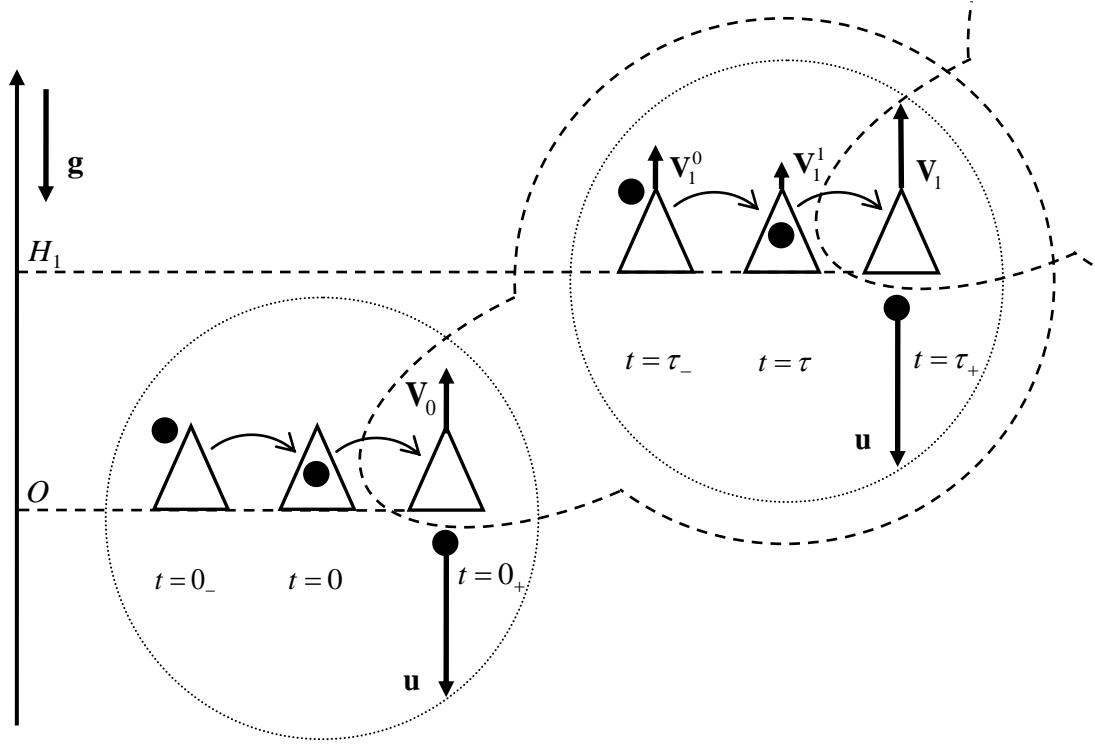

**Figure 1.** Cartoon for the pulse cycle model in the short pulses hypothesis. Velocities and positions are sketched for the first and second pulse. The triangle represents the engine (of mass *M*) and the black circle represents the fluid (of mass *m*) which merges with or is expelled by the engine. The magnitude and direction of vectors $\mathbf{V}_1^0$, $\mathbf{V}_1^1$, $\mathbf{V}_1$ are shown schematically for one of the possible cases; **g** is the gravitational vector field and **u** is the vector of the exhaust velocity (with respect to the engine). The dashed border domain suggests the 3-stage freefall-intake-exhaust processes; the first freefall-intake-exhaust process takes place in the time intervals, $(0_+, \tau_-]$, $(\tau_-, \tau]$, $(\tau, \tau_+]$, respectively.

*Freefall*. Between $t = 0_+$ and the moment just before the second intake (at $t = \tau_-$), we model the engine motion as being a freefall (with initial velocity $V_0$ in LF), such that the velocity at $t = \tau_-$ is given by,

$$V_1^0 = V_0 - g\tau. \qquad (3)$$

The sign of velocity $V_1^0$ is decided by the value of the parameters entering Eq. (3).

*Intake*. The engine merges with the fluid of mass *m* (which has zero velocity), so that the momentum conservation is written as,

$$MV_1^0 = (m+M)V_1^1, \qquad (4)$$

where $V_1^1$ is the velocity just after the second intake (at $t = \tau$) in the LF.

*Exhaust*. The mass *m* of fluid is ejected from the back of the engine at a (relative) exhaust velocity *u*, so that the momentum conservation is written as,

$$(m+M)V_1^1 = MV_1 + m(-u + V_1^1), \qquad (5)$$

The right-hand side of Eq. (5) is the system momentum just after the second exhaust (at $t = \tau_+$); the velocity $-u + V_1^1$, obtained by velocity composition, represents the expelled fluid velocity with respect to the LF. From Eqs. (3-5), one obtains that the velocity just after the second exhaust (or equivalently just after the second pulse) is given by,

$$V_1 = \frac{V_0 - g\tau}{1 + f} + fu, \tag{6}$$

Equation (4) assumes the fluid merges with the engine and before being expelled it contributes to reducing the speed of the engine. Generally, the validity of this assumption is decided by the constructive characteristics of the engine. For cephalopods, since the water is "captured, then released" use of Eq. (4) is fully justified. For air intake valved pulsejet engine [7-9], the existence of an intake space and combustion chamber makes application of Eq. (4) reliable. The position immediately after the second pulse is obtained from the freefall between $t = 0$ and $t = \tau$ as

$$H_1 = H_0 + V_0\tau - g\tau^2/2, \tag{7}$$

with $H_0 = 0$. The motion continues from position $H_1$ with the launch velocity $V_1$ at time $t = \tau_+$ and a new 3-stage process of freefall-intake-exhaust. Then, the motion is described with Eqs. (3-7) by replacing the subscript 0 by 1 and the subscript 1 by 2. Generalizing, the velocity (with respect to the LF) recurrence is given by,

$$V_n = \frac{V_{n-1} - g\tau}{1 + f} + fu, \quad V_0 = fu, \tag{8}$$

and the position recurrence is given by,

$$H_n = H_{n-1} + V_{n-1}\tau - \frac{g\tau^2}{2}, \quad H_0 = 0, \tag{9}$$

where $V_n$ and $H_n$ are the velocity and position just after the $(n+1)$th pulse. The pulse duration can easily be counted in the process, by inserting a new freefall between the intake and the exhaust moment. If the buoyancy is significant, the velocity rate of change during the freefall is lower. The solutions of the recurrences from Eqs. (8) and (9) are as follows:

$$V_n(\tau) = [f + C(n)]u - \frac{g\tau}{f}C(n), \tag{10}$$

$$H_n(\tau) = \frac{1+f}{f}[fn - C(n)]u\tau - \frac{g\tau^2}{2f^2}[fn(f+2) - 2(f+1)C(n)], \tag{11}$$

where $C(n) = 1 - (1+f)^{-n}$ and the pulse period dependence of the solutions is emphasized by the introduced $\tau$ argument.

The velocity and position as *continuous* functions of time can be obtained by taking intervals of the form $\text{Int}(t/\tau)\tau \leq t < [\text{Int}(t/\tau) + 1]\tau$, where $\text{Int}(t/\tau)$ is the integer part of the ratio $t/\tau$. Thus, we can write the velocity as

$$\upsilon(t) = V_{\text{Int}(t/\tau)}(\tau) - g\left[t - \text{Int}(t/\tau)\tau\right], \tag{12}$$

and the position as

$$h(t) = H_{\text{Int}(t/\tau)}(\tau) + V_{\text{Int}(t/\tau)}(\tau)\left[t - \text{Int}(t/\tau)\tau\right] - \frac{g\left[t - \text{Int}(t/\tau)\tau\right]^2}{2}. \tag{13}$$

Next, we discuss the characteristics of the motion as function of the parameters $\tau$, $f$, $u$, and $g$. *First*, to permanently have velocity $V_n$ directed against the gravitational field, we impose the condition: (*a*) $V_n(\tau) > 0$. *Second*, to have a permanently increasing velocity $V_n$, we impose the condition: (*b*) $V_{n+1}(\tau) > V_n(\tau)$. *Third*, to have not a return of the engine to the starting position, we impose the condition: (*c*) $H_n(\tau) > 0$. *Fourth*, to have a faster than linear increase of $H_n(\tau)$ with $n$, we require a positive second derivative of $H_n$ as function of $n$, that is: (*d*) $H_{n+2}(\tau) - 2H_{n+1}(\tau) + H_n(\tau) > 0$. The solutions of the inequations (*a*) – (*d*), which should be satisfied for any natural $n$, are presented in Table 1 (the second column shows an equivalent expression of the inequation from the first column). We find the limit values of the period $\tau$ for which the inequations are satisfied, in order, as follows: $\tau_{v1} = f(1+f)u/g$, $\tau_{v2} = fu/g$, $\tau_{h1} = 2f(1+f)u/[(2+f)g]$, $\tau_{h2} = fu/g$. Noticing the equality $\tau_{v2} = \tau_{h2}$, we increasingly re-order the period values and re-denote them as follows: $\tau_1(= \tau_{v2} = \tau_{h2}) < \tau_2(= \tau_{h1}) < \tau_3(= \tau_{v1})$.

**Table 1.** Calculus of the periods, $\tau_1$, $\tau_2$, $\tau_3$.

| Inequation | Equivalence of inequation | Solution of inequation |
|---|---|---|
| (*a*) | $\tau < f_1(n) = \frac{fu}{g}\frac{1+f-(1+f)^{-n}}{1-(1+f)^{-n}}$, $f_1(n)$ is decreasing with $n$ | $\tau < \lim_{n\to\infty} f_1(n) = \frac{f(1+f)u}{g} = \tau_3$ |
| (*b*) | $f_2(n) = (1+f)^{-n-1}(fu-g\tau) > 0$. | $\tau < \frac{fu}{g} = \tau_1$ |
| (*c*) | $\tau < f_3(n) = \frac{2fu}{g}$ $\times \frac{1}{2-f^2n(1+f)^{-1}\left[-1+(1+f)^{-n}+fn\right]}$, $f_3(n)$ is decreasing with $n$ | $\tau < \lim_{n\to\infty} f_3(n) = \frac{2(1+f)}{2+f}\frac{fu}{g} = \tau_2$ |
| (*d*) | $f_4(n) = (1+f)^{-n-1}\tau(fu-g\tau) > 0$ | $\tau < \frac{fu}{g} = \tau_1$ |

Searching for solution of $H_n(\tau) = 0$, we obtain the equation,

$$(1+f)^{-n} + cn = 1, \tag{14}$$

with $c = f\left[(2+f)g\tau - 2f(1+f)u\right]/\left[2(1+f)(gt-fu)\right]^{-1}$. By using the Lambert *W* function [10], the solution of the above equation can be written as,

$$n = \frac{\ln(1+f) + cW(b)}{c\ln(1+f)}, \quad (15)$$

where $b = -(1+f)^{-c}\ln(1+f)/c$. The integer part of the *real positive n* (that we denote by Int($n$)) plus one unit represents the number of pulses for the motion in which the engine returns to the starting position (after the Int($n$)+1 pulse and before the Int($n$)+2 pulse, the engine touches the starting position). Remarkably, we obtain that Int($n$) is *independent* of $u$ and $g$ if $\tau = \tau_3$ (since $c = f/2$), namely,

$$\text{Int}(n^*) = \frac{2}{f} + \frac{W\left(-2f^{-1}(1+f)^{-2/f}\ln(1+f)\right)}{\ln(1+f)}, \quad (16)$$

where we introduced the notation $\text{Int}(n)\big|_{\tau=\tau_3} \equiv \text{Int}(n^*)$, and ln is the natural logarithm function. To obtain the time the engine returns to the starting position, in Eq. (13) we replace $\text{Int}(t/\tau)$ by $\text{Int}(n^*)$ and take the larger root of the equation $h(t) = 0$ (in this case of real positive $n$, $h(t)$ is a polynomial of second degree in $t$, with real positive roots; the root of smaller value does not belong to the engine trajectory). The limit of $V_n$ is obtained from Eq. (10) as

$$V_\infty = \lim_{n\to\infty} V_n = (1+f)u - \frac{g\tau}{f}. \quad (17)$$

If the weight and buoyancy cancels each other (that is $g = 0$), the speed increases to the terminal speed value of $(1+f)u$. For $\tau < \tau_1$, $V_n$ is also asymptotically increasing, while for $\tau_1 < \tau \leq \tau_3$, $V_n$ is asymptotically decreasing to the value $V_\infty$ (for $\tau = \tau_3$, according to the definition from Table 1, first row, $V_\infty = 0$).

## 2.1 Comparison between pulsejet model and other models of reaction engines

Next, we compare the pulsejet with other two reaction engine models. First, considering the first freefall-intake-exhaust process (see Figure 1), the momentum of the two body system changes from $MV_0$ to $MV_1^0$ (at $t = 0_+$, and $t = \tau_-$, respectively, when the fluid amount $m$ is at rest in the environment). Then, as stated by Eqs. (4) and (5), the momentum remains constant and at $t = \tau_+$ it is written as $MV_1 + m(-u+V_1^1)$ (when the merged fluid amount $m$ is expelled). According to the impulse-momentum theorem, the momentum change of the two body system is equal to the product of the average external force and duration, that is,

$$MV_1^0 - MV_0 = MV_1 + m(-u+V_1^1) - MV_0 = -Mg\tau. \quad (18)$$

Notice that the first and last equality in Eq. (18) may also be thought as resulting from Eq. (3) multiplied by the engine mass $M$. By using Eqs. (3-5), Eq. (18) is re-written as,

$$MV_1 + m\left(-u + \frac{M}{m+M}V_0\right) - MV_0 = -Mg\tau\left(1 - \frac{m}{m+M}\right). \quad (19)$$

Second, we introduce the continuously operating turbojet model. The impulse-momentum theorem applied to the turbojet is written as,

$$(M - \Delta m_f)(V + \Delta V) + (\Delta m + \Delta m_f)(-u + V) - MV = -Mg\Delta t, \quad (20)$$

where $M$ is the aircraft plus fuel mass (at some moment), and $\Delta m$ and $\Delta m_f$ are the expelled small amounts of air and the fuel lost by the turbojet, respectively, in the time interval $\Delta t$. The left side term of Eq. (20) is the momentum variation (the first two terms represent the linear momentum at $t + \Delta t$ moment and the third term represents the linear momentum at the moment $t$, with respect to the LF). By neglecting the second order small variation, in the infinitesimal duration limit, from Eq. (20) one obtains the equation of motion [11],

$$M\dot{V} = q(u - V) + q_f u - Mg, \quad (21)$$

where $q = dm/dt$ is the air flow rate through the engine, and $q_f = dm_f/dt$ is the flow rate of the fuel lost by the turbojet. By also neglecting the tiny fuel momentum $\Delta m_f u$ in Eq. (20), one obtains,

$$M(V + \Delta V) + \Delta m(-u + V) - MV = -Mg\Delta t. \quad (22)$$

By comparing Eqs. (19) and (22), and noticing the correspondences, $V \leftrightarrow V_0$, $V_1 \leftrightarrow V + \Delta V$, $m \leftrightarrow \Delta m$, $\tau \leftrightarrow \Delta t$, one observes that Eqs. (19) and (22) have similar structure in the limit of small $f$. We conclude, that the pulsejet model is the discrete version of the turbojet model, with the difference that the intake-exhaust process duration, which is $\Delta t$ for turbojet, is negligible compared to the pulse period $\tau$ in the case of the pulsejet. In addition, the turbojet starts moving against gravity only if $q > Mg/u$, while the pulsejet has a starting velocity (against gravity) $V_0 (= fu)$.

With the initials conditions, $V(0) = 0$ for velocity and $H(0) = 0$ for position, by neglecting the fuel flow rate $q_f$ and by reasonably assuming that $q = m/\tau$, from Eq. (21) we obtain by integration,

$$V(t) = [u - g\tau/f]\left(1 - e^{-ft/\tau}\right) \quad (23)$$

for velocity, and

$$V_\infty = u - g\tau/f \quad (24)$$

for the terminal speed. By integrating Eq. (23), one obtains the position,

$$H(t) = \frac{fu - g\tau}{f^2}\left[ft + \tau(e^{-ft/\tau} - 1)\right]. \quad (25)$$

Equations (17) and (24) show similarity regarding the form of the terminal velocity of the two models.

We also analyze the dynamics of a virtual aircraft (we name it test aircraft) whose motion is described by a more severe approximation of Eq. (21), namely:

$$M\dot{V} = qu - Mg, \quad (26)$$

which is a simple one-dimensional motion with constant acceleration. With the initial conditions, $V(0) = 0$ for velocity and $H(0) = 0$ for position, by using $q = m/\tau$, we obtain the velocity,

$$V(t) = \left(\frac{fu}{\tau} - g\right) t, \tag{27}$$

and position,

$$H(t) = \left(\frac{fu}{\tau} - g\right) \frac{t^2}{2}. \tag{28}$$

## 3. Applications of the theory

In the usual pulsejet engines equipping the aircrafts, the compression of the air before mixing with fuel is insignificant. For the valved pulsejet engine Argus As 014 equipping the V-1 flying bomb, an estimated volume of the intake chamber of $0.5 \text{m}^3$ [9] contains approximately 0.64kg air at standard conditions of pressure and temperature. For the case of the VTOL aircraft, we consider the data from Table 2 by assuming $m$=1kg. In principle, larger air intake mass leads to increased thrust force. For a VTOL aircraft, compressing the air before its access to the combustion chamber may be a solution for increasing the engine efficiency [6]. Such an air compressor is a common component of the turbojet engines [11]. With the data from Table 2, the frequencies corresponding to three representative periods $\tau_1, \tau_2, \tau_3$ ($\tau_1 < \tau_2 < \tau_3$) are $\nu_1 = 98.00\text{s}^{-1}$, $\nu_2 = 97.97\text{s}^{-1}$, $\nu_3 = 97.95\text{s}^{-1}$, respectively; the values are in accordance with the real ones for the pulsejet engine [6]. With the data from Table 2 for the aircraft, the velocity and position just after the $(n+1)$th pulse, $V_n$ (with Eq. (10)) and $H_n$ (with Eq. (11)) as function of number of pulses are shown in Figure 2 for several representative values of the pulse period. The initial velocity is $V_0 = fu = 10^{-1}\text{m/s}$ and the initial position is $H_0 = 0$.

**Table 2.** Parameters of dynamics

| | $m$ [kg] | $M$ [kg] | $u$ [m/s] | $V_\infty$ [m/s] | $r$ [m] | $\tau$ [s] | $C_d$ | $\rho_f$ [kg/m$^3$] | $\eta$ [kg/(m·s)] |
|---|---|---|---|---|---|---|---|---|---|
| Aircraft | 1[a] | 2000[b] | 200[c] | 102.1[a] | 1[b] | 0.005[c] | 0.4[c] | 1.225[d] | 1.827×10$^{-5}$[e] |
| Octopus[f] *Sepia officinalis* | 0.0133 | 0.23 | 1.5 | 0.579 | 0.037 | 0.15 | 0.47[f] | 1025[g] | 1.08×10$^{-3}$[h] |
| Octopus[f] *Eledone moschata* | 0.2 | 0.6 | 9.4 | - | 0.06 | 10.58[*] | | | |

[a] Assumed values; [b] [9]; [c] [6]; [d] [12]; [e] [13]; [f] in accordance with [1, 2]; [g] [14]; [h] [15]; [*] $\tau = \tau_3$.

In Figure 2a, the velocity characteristics are shown. For $\tau = \tau_1$, $V_n$ is constantly equal to $V_0$. In this case, just after the first pulse the velocity is $V_0$, then the aircraft moves decelerating until the moment $\tau_1$ when its velocity becomes zero, and then the motion is repeated. For period longer than $\tau_1$, the velocity asymptotically decreases to the limit value, $V_\infty$. For $H_n$, the distinct cases are shown in Figure 2b. Thus, we obtain: for $\tau < \tau_1$, $H_n$ rapidly goes to infinity; for $\tau = \tau_1$, $H_n$ linearly increases with $n$ according to the law $f^2 n u^2 /(2g)$; for $\tau_1 < \tau < \tau_2$, $H_n$ increases with a slope smaller than $f^2 u^2 /(2g)$; for $\tau = \tau_2$, $H_n$ tends asymptotically to $2f(1+f)^2 u^2 /[(2+f)^2 g]$; for $\tau = \tau_3$ the aircraft returns to the starting position after $\text{Int}(n^*)+1 = 3187$ pulses (for $\tau > \tau_2$ the aircraft returns to the starting position).

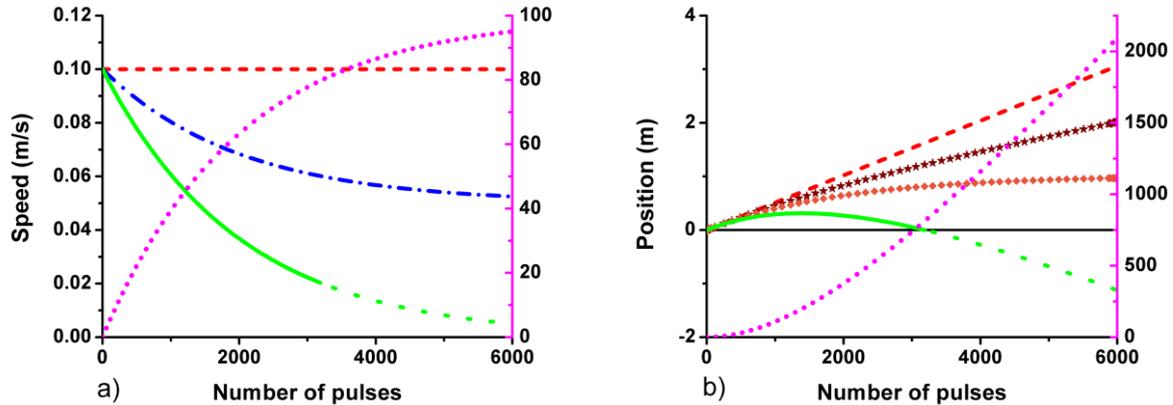

**Figure 2.** The discrete velocity $V_n$ (with Eq. (10)) and position $H_n$ (with Eq. (11)), just after the $(n+1)$th pulse as function of number of pulses: a) velocity for $\tau = \tau_1/2$ (magenta colour-dotted line), $\tau = \tau_1$ (red colour-dashed line), $\tau = (\tau_1+\tau_3)/2$ (blue colour-dashed dotted line), $\tau = \tau_3$ (green colour-continuous plus dashed line); b) position for $\tau = \tau_1/2$ (magenta colour-dotted line), $\tau = \tau_1$ (red colour-dashed line), $\tau = (\tau_1+\tau_2)/2$ (brown colour-star symbol), $\tau = \tau_2$ (orange colour-rhomb symbol), $\tau = \tau_3$ (green colour-continuous plus dashed line). The dashed continuation of the continuous (green colour) curves show the motion for the negative position coordinate in the case $\tau = \tau_3$. The right vertical axes (magenta colour) correspond to the case $\tau = \tau_1/2$ (magenta colour-dotted line). The initial velocity and position (just after the first pulse) have the values, $V_0 = fu$ and $H_0 = 0$, respectively. The parameters used are those from Table 2.

In Figure 3, the velocity and position for the pulsejet, turbojet, and test aircraft obtained with Eqs. (12), (23), (27), and (13), (25), (28), respectively, are presented for the data from Table 2. For the take-off of the turbojet and test aircraft, the initial acceleration should be positive, and according to Eq. (21) (with $q_f = 0$) or (26), respectively, the period should be shorter than $\tau_1$ (with $q = m/\tau$). For Figure 3, we chosen the period $\tau = \tau_1/2$. In Figure 3a, the velocity of the pulsejet and turbojet are of close values and both increase with time at a rate lower than that of the uniformly accelerated test aircraft. The characteristic 'saw tooth' shape of the velocity, which instantaneously increases after the pulse and then linearly decreases in the freefall, is shown in the inset of Figure 3a. Short time after the start, the turbojet and the test aircraft have almost identical velocities. This can analytically be obtained by noticing the equality between Eq. (27) and the series expansion for short time of $V(t)$ from Eq. (23). In Figure 3b, we show the variation with the time of the pulsejet and turbojet position; they are similar and have a slower increase than that parabolic of the uniformly accelerated test aircraft. In the inset of Figure 3b, one shows the characteristic parabolic shape of the position as function of time for the freefall motions of the pulsejet engine. Short time after the start, the turbojet and the test aircraft have almost identical variation of position with time. This can analytically be proved by noticing the equality between Eq. (28) and the series expansion of second order for short time of $H(t)$ from Eq. (25). For a longer time, the position coordinate of the test aircraft, which is a parabolic function, increases faster than the pulsejet or turbojet position. Crossing of the pulsejet and test aircraft positions can be obtained by equating Eqs. (13) and (28). The numerical solution for the crossing obtained with FindRoot of Mathematica is approximately 0.56s. The kinematics similarities of the pulsejet and turbojet, as concluded in section 2.1, are shown for velocity and position in Figure 3a and Figure 3b, respectively.

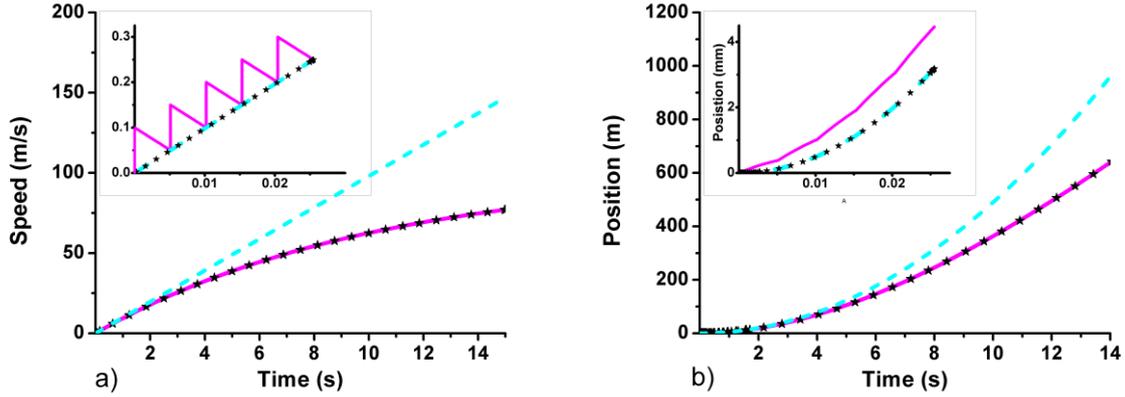

**Figure 3.** The continuous velocity and position for the aircraft models for the period $\tau = \tau_1/2$: a) $\upsilon(t)$ for pulsejet with Eq. (12) (magenta colour-continuous line), $V(t)$ for turbojet with Eq. (23) (black colour-star symbols), $V(t)$ for test aircraft with Eq. (27) (cyan colour-dashed line); b) $h(t)$ for pulsejet with Eq. (13) (magenta colour-continuous line), $H(t)$ for turbojet with Eq. (25) (black colour-star symbols), $H(t)$ for test aircraft with Eq. (28) (cyan colour-dashed line). The insets present details of motion at short time after the start. The parameters used are those from Table 2.

Next, we discuss the case when the engine returns to the starting position, and consider the period $\tau = \tau_3$. From Eq. (12), we obtain that just after the $(n+1)$th pulse the engine velocity is $\upsilon(n\tau_3 + 0_+) = V_n(\tau_3) = fu(1+f)^{-n} > 0$ (the engine is in ascending motion) and just before the $(n+2)$th pulse the engine velocity is $\upsilon(n\tau_3 + \tau_{3-}) = V_n(\tau_3) - g\tau_3 = -f\left[1 + f - (1+f)^{-n}\right] < 0$ (the engine is in descending motion). As an application, we estimate a fictitious periodic pulsejet motion of an octopus (*Eledone moschata*) with the data form Table 2 by estimating the octopus body density, $\rho_s$, as 4% greater than sea water [2]. If in the case of the aircraft dynamics model neglecting buoyancy is a good approximation, in the case of the octopus, buoyancy has important influence on the motion. Thus the gravity field intensity obtained as the resultant of weight and buoyancy (divided by mass), $g_0 = g(\rho_s - \rho_f)/\rho_s$, is approximately equal to 0.38m/s² for sea water density $\rho_f$ of $1025 \text{Kg/m}^3$ [14] and Earth's gravity field intensity of 9.8m/s². From refs. [1, 2], the pulse duration is in the range $0.15 \div 0.6\text{s}$ and it is negligible comparatively to the period $\tau_3 = 10.85\text{s}$. Consequently the assumption of short pulse is reasonable in this case. Velocity and position for this type of motion obtained with Eqs. (12) and (13), respectively, are shown in Figure 4a, where one assumes that after the octopus returns to the starting position, it remains at rest until the motion restarts with a new pulse. For arbitrary values of $u$ and $g$, we obtain $\text{Int}(n^*) = 4$. The $\text{Int}(n^*)$ independency of $u$ and $g$ is illustrated in Figure 4b for position as function of time.

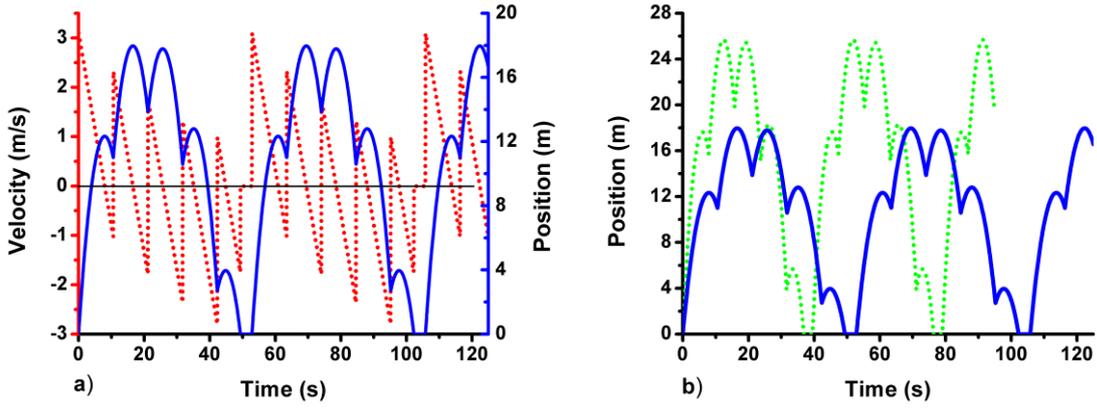

**Figure 4.** The continuous velocity $v(t)$ (with Eq. (12)) and position $h(t)$ (with Eq. (13)) for the octopus returning to the starting position: a) *Eledone moschata*, velocity (red colour-dotted line) and position (blue colour-continuous line) for $u = 9.4$m/s and $g_0 = 0.38$m/s$^2$ with the left vertical (red colour) axis for velocity and the right vertical (blue colour) axis for position; b) position with $u = 9.4$m/s and $g_0 = 0.38$m/s$^2$ (blue colour-continuous line) for *Eledone moschata*, and with $u = 18$m/s and $g_0 = 1$m/s$^2$ (green colour-dotted line) for a fictional octopus in fictional gravitational field. For both figures a) and b), $f = 0.(3)$.

The assumptions (i)-(iii) of the frictionless model from Introduction may be considered as reasonable approximations even for more realistic models. On another hand, assumption (iv) can be relaxed and the pulse duration may easily be introduced into the model with a necessary re-evaluation of the conclusions. Assumption (v) is the roughest approximation and next we shortly discuss its validity. For lower velocity the viscous forces become dominant over the inertial forces, the fluid flow around the object is laminar and the friction is a linearly velocity dependent drag force. For higher velocity the forces reverse their magnitude, the fluid flow around the object becomes turbulent, and the friction becomes a quadratic velocity dependent drag force. To quantitatively describe the flow type, one introduces the Reynolds number (defined as the ratio of inertial forces to viscous forces [16]):

$$Re = \rho_f L V / \eta \qquad (29)$$

where $\rho_f$ is density of the fluid, $L$ is a characteristic linear dimension of the object, $V$ is the velocity of the object relative to the fluid, and $\eta$ is the fluid viscosity. A laminar flow has low Reynolds numbers, while the turbulent flow is characterized by high Reynolds numbers. For example, the critical Reynolds number which characterizes transition from laminar to turbulent flow for a sphere is about 100 [17]. With the data from Table 2, by approximating the aircraft and octopus as being of spherical shape and considering a circular cross-section of maximum diameter $L = 2r$, for the tenth part of the terminal speed, $V_\infty$, we obtain the Reynolds number $Re_{VTOL} = 1.37 \times 10^6$ for the aircraft and $Re_{oct} = 4.15 \times 10^3$ for the *Sepia officinalis*, thus in both cases the flow is turbulent. Though considering the drag forces dependency on velocity makes the problem complex, we can estimate the frictionless model capability of explaining the real motion. Thus, if the thrust force is much stronger than the drag force, the estimation becomes more reasonable. For the pulsejet, the *average net* thrust varies from the value in the first freefall-intake-exhaust process (see Eq. (18)),

$$F_N^1 = \frac{m}{\tau}(u - V_1^1) = \frac{m}{\tau}\left(u - \frac{MV_0}{m+M}\right) + \frac{mM}{m+M}g, \qquad (30)$$

to the value in the *n*th freefall-intake-exhaust process

$$F_N^n = \frac{m}{\tau}(u - V_n^1) = \frac{m}{\tau}\left(u - \frac{MV_{n-1}}{m+M}\right) + \frac{mM}{m+M}g, \qquad (31)$$

which for infinite *n* becomes $Mg$ (see Eq. (17)). For example, for $\tau = \tau_1/2$, one obtains $F_N^n = \left[1 + (1+f)^{-n}\right]Mg$, and the velocity progresses with the saw tooth shape shown in Figure 3a. Similarly, for the turbojet, the average net thrust

$$F_N = m(u - V)/\tau, \qquad (32)$$

varies from $mu/\tau$ at the initial moment to $Mg$ at infinity (see Eq. (24)). According to Eq. (32), at infinity, the average thrust $F_N - Mg \to 0$ and this explains the existence of the constant terminal velocity as given by Eq. (24). Similarly, according to Eq. (31), at infinity, the pulsejet average thrust $F_N^n - Mg \to 0$ and this explains the existence of a pulse terminal velocity as given by Eq. (17). With the data from Table 2, in Figure 5, we represent the pulsejet average thrust

$$F^n = F_N^n - Mg, \qquad (33)$$

(as function of time $t = n\tau$) and the corresponding quadratic drag force [11]

$$F_d^n = \rho_f V_n C_d A/2, \qquad (34)$$

(*A* is the cross-sectional area of the moving system) to estimate the validity of the drag-free approximation.

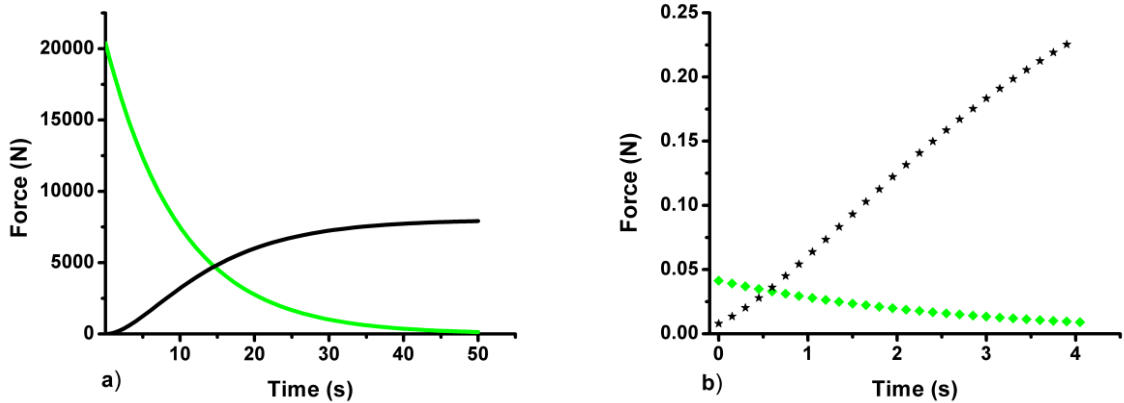

**Figure 5.** The average thrust $F^n$ (with Eq. (33)) and drag force $F_d^n$ (with Eq. (34)) as function of time $t = n\tau$ for: a) pulsejet aircraft, thrust (green colour - grey line) and drag force (black line); b) *Sepia officinalis*, thrust (green rhomb symbol - rhomb symbol) and drag force (star symbol), with the data from Table 2. The high frequency of the pulses makes the discrete character of the graph not visible in Figure 5a. One observes that $F^n \to 0$ as the time $t = n\tau$ increases.

Generally, as a criterion, the drag-free approximation is reliable as long as $F_N^n - Mg \gg F_d^n$, which happens for a lower speed of the engine. Figure 5a shows that during the first 5s after the launch, the pulsejet thrust force is at least ten times stronger than the drag force (5s after the launch, $V_n = 40.2$m/s). The same ordering relation between the two forces is obtained by taking $m=0.55$kg in the data from Table 2 (one has $f = 275 \times 10^{-6}$ and $V_\infty = 21.87$m/s) during the first 15s after the launch (15s after the launch, $V_n = 12.31$m/s). The graphs of the thrust and quadratic drag force for the turbojet with the data $M$, $\tau$, and air flow rate $q = m/\tau$ for the pulsejet from Table 2, practically superpose on those of the pulsejet (not shown). Figure 5b, which is for *Sepia officinalis*, shows that after the fourth pulse the drag force already overcomes the thrust force. Thus the validity criterion of the drag-free approximation is broken more quickly with the number of pulses in the octopus motion case.

## 4. Conclusions

In conclusion, by applying the momentum conservation we successfully modelled the frictionless vertical dynamics of the pulsejet engine in a uniform gravitational field. The modelling reveals typical kinds of motion of the pulsejet engine and some interesting characteristics. Thus in the vertical motion, we obtain: i) the engine velocity (with respect to the LF) has the upper limit $(1+f)u$; ii) for period shorter (longer) than $\tau_1 = fu/g$, the velocity asymptotically increases (decreases) to the terminal value $V_\infty = (1+f)u - g\tau/f$); iii) for period longer than $\tau_2 (= 2f(1+f)u/[(2+f)g])$ the engine returns to the starting position; iv) for period equal to $\tau_3 (= f(1+f)u/g)$ there is an independent of $u$ and $g$ number of short periodic pulses after which the engine returns to the starting position (the relevant periods are ordered as follows, $\tau_1 < \tau_2 < \tau_3$). Comparison between the pulsejet and turbojet dynamics models shows similarities of the velocity, position, and average thrust force as function of time of the two models. Regarding the modelling accuracy, we conclude: (i) the estimation of real vertical motion against gravity by the frictionless model is more reliable for speeds generally much lower than the terminal speed, and (ii) the reliability of modelling is higher for motion in air (the aircraft case) than for motion in water (the cephalopods case).

Regarding the educational relevance, the power of the physics conservation laws in modelling the reality is exemplified. The modelling of the pulse engine as a pair of two perfect inelastic collision processes and explanation of the propulsion by the momentum conservation is pedagogically considered. With a correct understanding of the momentum conservation in the functioning of the reaction thrust engine and basic knowledge of mathematical analysis, interesting physics is found. A complete description of the frictionless dynamics allows one to find an exhaust velocity and gravity independent parameter of motion (the integer number $\text{Int}(n^*)$). The modelling accuracy of the frictionless model we introduced may be improved by considering more accurately the velocity dependency of the drag forces.